# Quantum spin liquid states in the two dimensional kagomé antiferromagnets, $Zn_xCu_{4-x}(OD)_6Cl_2$

S.-H. Lee[*], H. Kikuchi[†], Y. Qiu[‡], B. Lake[$,+], Q. Huang[‡], K. Habicht[$] and K. Kiefer[$]

[*] Department of Physics, University of Virginia, Charlottesville, Virginia 22904-4714, USA

[†] Department of Applied Physics, University of Fukui, Fukui 910-8507, Japan

[‡] NIST Center for Neutron Research, National Institute of Standards and Technology, Gaithersburg, Maryland 20899, USA

[$] Hahn-Meitner-Institut, Glienicker Strabe 100, Berlin D-14109, Germany

[+] Institut für Festkörperphysik, Technische Universität Berlin, Hardenbergstr. 36, 10623 Berlin, Germany

A three-dimensional system of interacting spins typically develops static long-range order when it is cooled. If the spins are quantum (S = 1/2), however, novel quantum paramagnetic states may appear. The most highly sought state among them is the resonating valence bond (RVB) state[1,2] in which every pair of neighboring quantum spins form entangled spin singlets $\left( \left| \uparrow \downarrow \right\rangle - \left| \downarrow \uparrow \right\rangle \right)$ (valence bonds) and the singlets are quantum mechanically resonating amongst all the possible highly degenerate pairing states. Here we provide experimental evidence for such quantum paramagnetic states existing in frustrated antiferromagnets, $Zn_xCu_{4-x}(OD)_6Cl_2$, where the S = 1/2 magnetic $Cu^{2+}$ moments form layers of a two-dimensional kagomé lattice. We find that in $Cu_4(OD)_6Cl_2$, where distorted kagomé planes are weakly coupled to each other, a dispersionless excitation mode appears in the magnetic excitation spectrum below ~ 20 K, whose characteristics resemble those of quantum spin singlets in a solid state, known as a valence bond solid (VBS), that breaks translational symmetry. Doping nonmagnetic $Zn^{2+}$ ions reduces the distortion of the kagomé lattice, and weakens the interplane coupling but also dilutes the magnetic occupancy of the kagomé lattice. The VBS state is suppressed and for $ZnCu_3(OD)_6Cl_2$ where the kagomé planes are undistorted and 90% occupied by the $Cu^{2+}$ ions, the low energy spin fluctuations in the spin liquid phase become featureless.

The RVB state was originally introduced by Anderson in the early 1970s as a possible ground state for two-dimensional triangular quantum antiferromagnets[1,2]. A decade later, Anderson proposed the formation of the RVB state as a mechanism for high $T_c$ superconductivity[3]. Since then, such quantum spin liquid (QSL) states have been greatly sought after in various strongly correlated electron systems. Theoretically, substantial progress has been made on the physics of quantum paramagnets[4-10]. For instance, it has been shown that there exists another simpler type of cooperative quantum state known as a valence bond solid (VBS) in which each spin forms a spin singlet $\left( \left| \uparrow \downarrow \right\rangle - \left| \downarrow \uparrow \right\rangle \right)$ or (valence bond) with a neighbor, resulting in an ordered pattern (solid) of valence bonds. It is not easy, however, to find QSLs in real systems because these states lack static order and their dynamic order parameters are hard to observe: for the VBS state, these are the gapped S = 1 quasiparticle excitations (triplons), and for the RVB state, these are an exotic gapless continuum of excitations with fractional spin S = 1/2 (deconfined spinons). Such states in systems with spatial dimensions larger than one have so far eluded all experimental attempts. Over the last two decades or so, frustrated magnets have been studied as excellent systems to look for such QSLs because of their intrinsic macroscopic ground state degeneracy[11,12].

Recently $Zn_xCu_{4-x}(OD)_6Cl_2$ has been proposed as an ideal candidate for the quantum (S = 1/2) kagomé system where quantum spins form a two-dimensional network of corner-sharing triangles[13-16] (see Fig. 1 (c)). We have performed elastic and inelastic neutron scattering measurements on this compound with various Zn concentrations (x = 0, 0.2, 0.4, 0.66, 1.0) and studied how the static and dynamic spin correlations evolve with x and temperature. For the x = 0, $Cu_4(OD)_6Cl_2$, where the kagomé layers are weakly coupled via $Cu^{2+}$ ions at the triangular sites located in between the $Cu^{2+}$ kagomé layers, our data shows two nearly dispersionless modes of magnetic excitations at $\hbar\omega_1 = 1.3$ meV and at $\hbar\omega_0 = 7$ meV, respectively (see Fig. S1 and S2 in the supplementary data). This system undergoes magnetic long range (Néel) order. Fig. 2 (a) shows that magnetic Bragg peaks develop below $T_N \sim 7$ K with a characteristic wave vector of (010) with respect to the $P2_1/n$ crystal symmetry[17]. This transition is consistent with the previously reported bulk susceptibility data[17,18]. The magnetic structure, refined using the elastic data, is shown in Fig. 1 (c) (for details, see the caption of Fig. 2 (a) and the supplementary information). Interestingly, the spins are not $120^o$ apart from each other as expected in the so-called q = 0 or $\sqrt{3} \times \sqrt{3}$ spin structures proposed as the possible ground states of a kagomé antiferromagnet. Instead, they are *collinear*, the implication of which will be discussed later. As the static order sets in, upon cooling, the low energy spectrum below 2.5 meV starts to develop a gap (Fig. 2 (b)) and the gap, $\hbar\omega_1$, increases as the static moment $<M>$ increases. It is well known that a Néel state with a single ion anisotropy develops a gap in some long ranged spin wave excitation modes[19]. For example, in the q = 0 kagomé Néel state, a zero-energy mode that involves swinging of spins about a bisecting axis of symmetry can be lifted in energy by an anisotropy[20,21]. We conclude that the $\hbar\omega_1$ mode is due to such nearly dispersionless spin wave excitations allowed by the Néel state of $Cu_4(OD)_6Cl_2$. This explanation is supported by the doping effect of nonmagnetic $Zn^{2+}$ ions. The Neel order weakens upon Zn doping, and disappears somewhere between x = 0.4 and 0.66 (Fig. 3 (a)). The $\hbar\omega_1$ mode is present as long as the Neel order is present (Fig. 3 (b)), but it shifts to lower energies as the ordered moment decreases. In a similar way the $\hbar\omega_1$ decreases as temperature increases in the x = 0 sample (Fig. 2 (b)). This is because the ordered moment decreases with $T$ ( $< T_N$) (see Fig. 2 (a)).

The $\hbar\omega_0 = 7.2$ meV mode of the x = 0 system, in contrast to the $\hbar\omega_1$ mode, exists up to $T_c \sim 20$ K which is well above its Néel phase ($T_N = 7$ K) (red circles in Fig. 2 (a)). Indeed previous specific heat, bulk susceptibility studies[17] and muon spin resonance measurements[16] reported a transition at ∼ 18 K. This confirms that the development of the $\hbar\omega_0$ mode below $T_c$ represents a phase transition between two different states. The question that arises is: what kind of states are they? The value of $\hbar\omega_0$ does not change below $T_c$ (the right hand side of Fig. 2 (b)), which starkly contrasts with the behavior of the $\hbar\omega_1$ mode below $T_N$. In addition, the Q-dependence of the $\hbar\omega_0$ mode has a peak at 1.1 $\overset{\circ}{A}^{-1}$, that is different from that of the $\hbar\omega_1$ mode which has a peak at ∼ 0.6 $\overset{\circ}{A}^{-1}$ (Fig. 4 (a)). These clearly tell us that the $\hbar\omega_0$ mode is not related to any spin wave of the Néel state.

The origin of the $\hbar\omega_0 = 7.2$ meV mode can be elucidated by its Q-dependence. As shown in Fig. 4 (a), the Q-dependence of the $\hbar\omega_0$ mode (red circles) resembles that of the singlet-to-triplet excitations of spin dimers (red line), $1 - \dfrac{\sin(Qr_0)}{Qr_0}$, [22,23] with $r_0 = $

3.41 $\overset{\circ}{\mathrm{A}}$, which is the distance between the nearest neighbor $Cu^{2+}$ (S = 1/2) ions in the

kagomé lattice. Furthermore, the mode energy $\hbar\omega_0 = 7.2$ meV is close to the estimated

coupling constant for $Cu_4(OD)_6Cl_2$, $J \sim 9.7$ meV from bulk susceptibility data[24]. These

strongly indicate that the $\hbar\omega_0$ mode is the gapped S = 1 quasiparticle excitation that is

the hallmark of a valence bond solid (VBS) state. We believe that the VBS state arises

due to the lattice distortion which results in broken translational symmetry and non-

uniform interactions in the kagomé plane. The valence bond singlets form around the

dominant interactions and the excitations correspond to promoting one of these into an

S = 1 triplet. Upon further cooling, the spins freeze into a collinear Néel state rather

than the noncollinear arrangements (e.g. q = 0 or $\sqrt{3} \times \sqrt{3}$ ) commonly found in

frustrated kagomé antiferromagnets. This ordering retains the antiparallel spin

alignment within the original singlets established at higher temperatures in the VBS

state. Above $T_c$, the $\hbar\omega_0$ mode disappears and is replaced by a featureless energy

continuum. Interestingly, however, the Q-dependence of the low energy continuum

above $T_c$ (black circles in Fig. 4(a)) is the same as that of the $\hbar\omega_0$ mode. In other

words, the spin liquid phase above $T_c$ has energy continuum excitations with the Q-

dependence of spin dimers, identifying this as a valence bond liquid (VBL). Upon

doping with the nonmagnetic $Zn^{2+}$ ions, $T_c$ decreases and the $\hbar\omega_0$ mode weakens, and

for x = 0.4 it disappears altogether (see Fig. 3 (b)). It is to be noted that the crystal

symmetry changes at x = 0.33 from monoclinic $P2_1/n$ (x < 0.33) to rhombohedral

R$\bar{3}$m (x > 0.33),[13] which coincides with the disappearance of the VBS state and the dramatic weakening of the Néel order.

For x ≥ 0.66 when there is no Néel order, the low energy excitations become gapless and also weaken considerably, which makes the investigation non-trivial. To observe any increase in low energy spin fluctuations upon cooling for x = 0.66, we subtracted the data measured at 10 K from the 1.5 K data. The difference, $\int_{0.2meV}^{0.8meV}$ I(1.5K) - I(10K) $d(\hbar\omega)$, (blue squares in Fig. 4 (b)) shows a broad peak centered at Q ~ 0.8 $\overset{o}{A}$ $^{-1}$, halfway between the two magnetic (001) and (010) Bragg peak positions of the x = 0 system (see Fig. 3 (a)), suggesting that this intensity is due to short range critical fluctuations around the Néel state. The low energy excitations grow below $T_f$ ~ 5 K (blue squares in the inset of Fig. 4) at which bulk susceptibility data (magenta lines) exhibit field-cooled and zero-field-cooled hysteresis, indicating a spin glass-like phase transition. When x = 1, on the other hand, not even spin freezing occurs at low temperatures. Only a nearly Q-independent increase is observed upon cooling at low temperatures (blue squares in Fig. 4(c)). Interestingly, the Q-dependence of low energy magnetic excitations (cyan squares) can be modelled by the squared magnetic form factor of the single $Cu^{2+}$ ion with S = 1/2 (cyan line). This behavior, characteristic of the excitations of uncoupled spins, is also observed in the magnetic field-induced excitations of the spin liquid phase of $Zn_xCu_{4-x}(OD)_6Cl_2$ with x ≥ 0.66. As shown in Fig. 5, when the field, $H$, is applied to the x = 0.66 sample, the energy continuum of magnetic excitations changes to develop an excitation mode at finite energies. The peak energy of the field-induced excitation is proportional to $H$: $E$ (meV) = 0.12(3) $H$ (Tesla) $\approx g\mu_B H$ where $g$ is the gyromagnetic ratio and $\mu_B$ is the Bohr magneton, and which is the expected Zeeman splitting behavior of a single magnetic ion in the presence of an external magnetic field. This result is consistent with the previously observed field effect on the x = 1 sample[14].

What is the origin of the single ionic excitations in the spin liquid phase of $Zn_xCu_{4-x}(OD)_6Cl_2$ with x ≥ 0.66? Spin liquid states that have been theoretically proposed for the perfect quantum kagomé antiferromagnet have strong characteristic Q-dependences due to strong spin correlations, which cannot explain the observed single ionic behavior. To understand the observed behavior let us consider the effect of Zn doping on the crystal structure. It has been assumed that since the $Zn^{2+}$ ions are not Jahn-Teller active, they prefer the triangular sites that are surrounded octahedrally by six $O^{2-}$ ions over the kagomé sites that are surrounded non-octahedrally by four $O^{2-}$ and two $Cl^{2-}$ ions (see Fig. 1 (a) and (b)). If the nonmagnetic $Zn^{2+}$ ions replace the $Cu^{2+}$ ions at the triangular sites only, but leave the $Cu^{2+}$ ions at the kagomé sites intact, $ZnCu_3(OD)_6Cl_2$ would be the fully occupied perfect two-dimensional quantum spin kagomé system. In order to find out if this is indeed the case, we have performed neutron powder diffraction measurements on the x = 1 sample and refined the crystal structure. Our results show that $Zn^{2+}$ ions prefer the octahedral triangular sites but they can also go into the non-octahedral kagomé sites. For x = 1 the triangular sites are 36% occupied by $Cu^{2+}$ ions and 64% by $Zn^{2+}$ ions while the kagomé sites are 90% occupied by $Cu^{2+}$ ions and 10% by $Zn^{2+}$ ions (see the supplementary information). The dilution of the kagomé lattice may provide a possible explanation for the single ionic spin fluctuations.

To test this scenario we have normalized our inelastic neutron scattering data to an absolute unit by comparing the magnetic intensity to the intensities of nuclear peaks[29] (see Figs. 3 (b) and 5). For x = 0 and at 1.5 K, the integrated intensity of the $\hbar\omega_1$ : 1.3 meV mode, $\int_{1 meV}^{1.7 meV} I(\hbar\omega) d(\hbar\omega) = 0.20(2)/Cu^{2+}$ that corresponds to 27 % of the total magnetic scattering cross section, S(S+1) = 0.75/Cu$^{2+}$. For x = 0.66 and 1, the integrated intensity over low energies up to 1.7 meV is $\int_{0.2 meV}^{1.7 meV} I(\hbar\omega) d(\hbar\omega) = 0.11(3)/Cu^{2+}$ and $0.13(3)/Cu^{2+}$, respectively, assuming that all Cu$^{2+}$ ions are involved. If we assume that only the unpaired triangular Cu$^{2+}$ ions are involved, the integrated intensity for x = 1 becomes 1.1(2)/Cu$^{2+}$ that is larger than S(S+1)/Cu$^{2+}$ and thus unrealistic. This indicates that the single-ionic low energy excitations observed in the spin liquid phase of the x = 0.66 and 1 are indeed due to collective excitations of kagomé Cu$^{2+}$ ions. It is possible that the VBL state observed for x = 0 evolves, upon increasing x, into a similar spin liquid state but with much shorter-range spin correlations that results in the featureless Q-dependence for x $\geq$ 0.66. Fully understanding the Q-dependence and the Zeeman response to the external magnetic field requires further theoretical studies on quantum kagomé antiferromagnets.

Despite the intense theoretical interest in the nature of spin fluctuations in a quantum kagomé system[5-10,28], experimental studies have been rare due to the scarcity of the good model systems. In this paper, we have presented our studies on Zn$_x$Cu$_{4-x}$(OD)$_6$Cl$_2$ that realizes a weakly coupled, spin-1/2, kagomé antiferromagnet, performed using elastic and inelastic neutron scattering with and without an external magnetic field. Our systematic studies of magnetic excitations in Zn$_x$Cu$_{4-x}$(OD)$_6$Cl$_2$ with various x, allow us to construct a schematic phase diagram as a function of x and temperature (Fig. 1 (d)). Various magnetic phases, such as Néel, valence bond solid, valence bond liquid and cooperative paramagnetic phases, were identified in different regions of the phase diagram, and the characteristics of their spin fluctuations were investigated. Full understanding of our results requires further theoretical and experimental studies on the quantum kagomé system, and the experimental results presented here would serve a crucial test for any theoretical models.

## Methods

We have performed elastic and inelastic neutron scattering measurements without field on 100% deuterated powder samples of $Zn_xCu_{4-x}(OD)_6Cl_2$ with five different Zn concentrations, x = 0, 0.2, 0.4, 0.66 and 1. The measurements were done at the NIST Center for Neutron Research using the time-of-flight Disk-Chopper-Spectrometer (DCS), the cold-neutron triple-axis spectrometer SPINS and the thermal triple-axis spectrometer, BT9. Experimental configurations that have basically two different energy resolutions were used for our measurements: one set of data was taken up to $\hbar\omega \sim 2.5$ meV with an energy resolution of $\Delta\hbar\omega \sim 0.1$ meV and the other set of data was taken up to $\hbar\omega \sim 15$ meV with a coarser $\Delta\hbar\omega \sim 1 - 2$ meV. We have also performed the measurements with an external magnetic field on $Zn_{0.66}Cu_{3.34}(OD)_6Cl_2$ at the Hahn-Meitner Institut using the triple-axis spectrometer, V2. The inelastic data shown in Fig. 2 (b) were obtained from the DCS raw data with empty can background subtracted. The data still contains nonmagnetic signal, especially at very low energies due to the instrumental energy lineshape that is a characteristic of a time-of-flight spectrometer. The magnetic and the nonmagnetic component were separated using the detailed balance relation $I(-\hbar\omega, T) = I(\hbar\omega, T) \exp(-\hbar\omega / k_B T)$ [29] and the magnetic components were plotted in Fig. 3 (b).

**Acknowledgements**

We thank D. Khomskii, S. Sachdev, and M. Gingras for helpful discussions. S.-H.L is partially supported by NIST. Activities at SPINS and DCS were partially supported by NSF.

Correspondence and requests for materials should be addressed to S.H.L. (e-mail: shlee@virginia.edu).

**Figure captions**

**Fig. 1.** Local crystal structure and phase diagram of $Zn_xCu_{4-x}(OD)_6Cl_2$. (a) Local environments around the nonmagnetic $Zn^{2+}$ ion (grey spheres) at the triangular site and the magnetic $Cu^{2+}$ ions (blue spheres) at the neighboring kagomé sites. The kagomé $Cu^{2+}$ ions are surrounded by four $O^{2-}$ (red spheres) and two $Cl^-$ ions in a distorted octahedral environment with an elongation along the $Cl^-$ axis. Here, for simplicity, only $O^{2-}$ ions are shown. The $Zn^{2+}$ ion is surrounded by six $O^{2-}$ ions in a perfect octahedral environment. (b) Local oxygen environments around the triangular site when the $Zn^{2+}$ ion is replaced by a magnetic $Cu^{2+}$ ion. Since a $Cu^{2+}$ ion is Jahn-Teller active, its oxygen octahedron distorts. The $Cu^{2+}$-$O^{2-}$ bonds where the wave functions of their unpaired electrons overlap are shown by grey bars. The superexchange interaction between nearest neighboring $Cu^{2+}$ moments mediated via oxygen is very sensitive to the Cu-O-Cu bond angle, $\theta$, and it changes from strongly antiferromagnetic when $\theta = 180^o$ to zero when $\theta \approx 95^o$ to ferromagnetic when $\theta < 95^o$ [26,27]. These angles are shown in the figure and indicate that interactions between the kagomé $Cu^{2+}$ and the doped triangular $Cu^{2+}$ spins are weak, which makes the $Cu^{2+}$ moments at the triangular sites almost independent spins and thus makes $Zn_xCu_{4-x}(OD)_6Cl_2$

a weakly coupled kagome system for all Zn concentration x. (c) The spin structure of $Cu_4(OD)_6Cl_2$ determined by fitting our elastic neutron scattering data shown in the inset of Fig. 2 (a) to a model of kagomé spins only (see the caption of Fig. 2 (a) for the details). The spins are collinear with each other: the up and down arrows represent antiparallel spins in the Néel state $T < T_N = 7$ K. The kagomé lattice is slightly distorted in $Cu_4(OD)_6Cl_2$ to have two bond lengths, 3.41 Å and 3.42 Å. The neighboring spins with the shorter bond length are antiparallel whereas those with the longer bond length are parallel. The dashed lines represent the chemical and magnetic unit cell. The ovals represent the quantum singlets $\left( \left| \uparrow \downarrow \right\rangle - \left| \downarrow \uparrow \right\rangle \right)$ in the valence bond solid (VBS) state of $Cu_4(OD)_6Cl_2$, $T < T_c = 20$ K. (d) Phase diagram of $Zn_xCu_{4-x}(OD)_6Cl_2$ as a function of temperature and doping.

**Fig. 2.** Energy and temperature dependence of the static and dynamic spin correlations in $Cu_4(OD)_6Cl_2$ (a) The inset shows elastic neutron scattering patterns obtained from a powder sample at 1.5 K (blue plus symbols) and 10 K (black line). Magnetic Bragg peaks, such as the (001) and (010) reflections, exist at 1.5 K, indicating the development of Néel order at low temperatures. The magnetic Bragg peaks could be reproduced by considering kagomé spins only with the ordered moment, <M>$_{kagomé}$, of $1.15 \pm 0.15 \mu_B / Cu^{2+}$. The data could also be fit equally well by considering both kagomé and triangular spins with <M>$_{kagomé} = 1.07 \pm 0.20 \mu_B / Cu^{2+}$ and <M>$_{triangular} = 0.59 \pm 0.45 \mu_B / Cu^{2+}$, respectively. The fact that the kagomé spins are sufficient to reproduce the data and that <M>$_{triangular} \sim 0.5$ <M>$_{kagomé}$ when the triangular spins are included supports our speculation from the crystal structure that the interactions between the kagomé and triangular spins are very weak. The blue line in this inset and the collinear spin structure shown in Fig. 1 (c) are the results obtained with the model of the kagomé spins only. The main figure (a) shows the detailed temperature dependence of <M>$_{kagomé}$ (blue squares) and that of the $\hbar\omega_0 = 7.2$ meV mode intensity (red circles). (b) Magnetic excitation spectrum of $Cu_4(OD)_6Cl_2$ at several different temperatures.

**Fig. 3.** Doping dependence of static and dynamic spin correlations. (a) Elastic neutron scattering intensities at two magnetic reflections obtained for $Zn_xCu_{4-x}(OD)_6Cl_2$ with x = 0, 0.2, 0.4, 0.66 and 1.0, and at 1.5 K., (b) Magnetic excitation spectrum in $Zn_xCu_{4-x}(OD)_6Cl_2$ with x = 0, 0.2, 0.4, 0.66, and 1.0, measured at 1.5 K.

**Fig. 4.** Q-dependence of magnetic fluctuations for (a) $Cu_4(OD)_6Cl_2$, (b) $Zn_{0.66}Cu_{3.34}(OD)_6Cl_2$, and (c) $ZnCu_3(OD)_6Cl_2$. Different colors represent the data obtained for different energies at different temperatures. In (a), black symbols represent I (T = 40 K, $h\omega = 4$ meV), red: I (T = 1.5 K, $h\omega_0$), and blue: I (T = 40 K, $h\omega_1$). (b) $\int_{0.2meV}^{0.8meV} I(1.5K) - I(10K) d(h\omega)$ (blue squares) obtained from the x = 0.66 sample that shows an increase of intensity at low energies below 10 K. The inset of (b) shows the temperature dependence of these low energy excitations (blue squares) and that of the bulk susceptibility (pink lines) measured under field-cooled (FC) and zero-field-cooled (ZFC) conditions, indicating spin-freezing below 5 K. (c) Low energy magnetic excitations measured at 30 K and with $h\omega = -0.6$ meV (cyan squares). Here I(1.5K) with the negative energy transfer was measured to determine nonmagnetic contributions using the principle of detailed balance, $I(-\omega) = \exp(-h\omega / k_B T) I(\omega)$ where $k_B$ is the Boltzmann factor, and was subtracted from I(30K). Blue squares are the difference in intensity between 1.5 K and 30 K, measured with $h\omega = 0.6$ meV.

**Fig. 5.** Magnetic field effect on magnetic excitations. Magnetic excitation spectrum observed from a powder sample of $Zn_{0.66}Cu_{3.34}(OD)_6Cl_2$ with application of various external magnetic fields, H = 0T, 7T, 12T and 13.5T. The inset shows the peak energy of the field-induced excitations as a function of H.

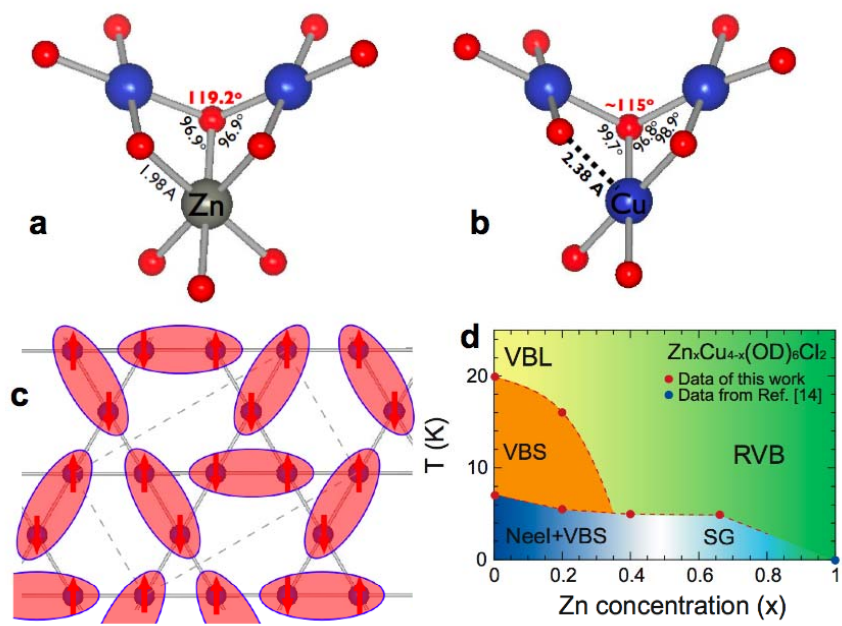

Fig. 1. S.-H. Lee *et al*.

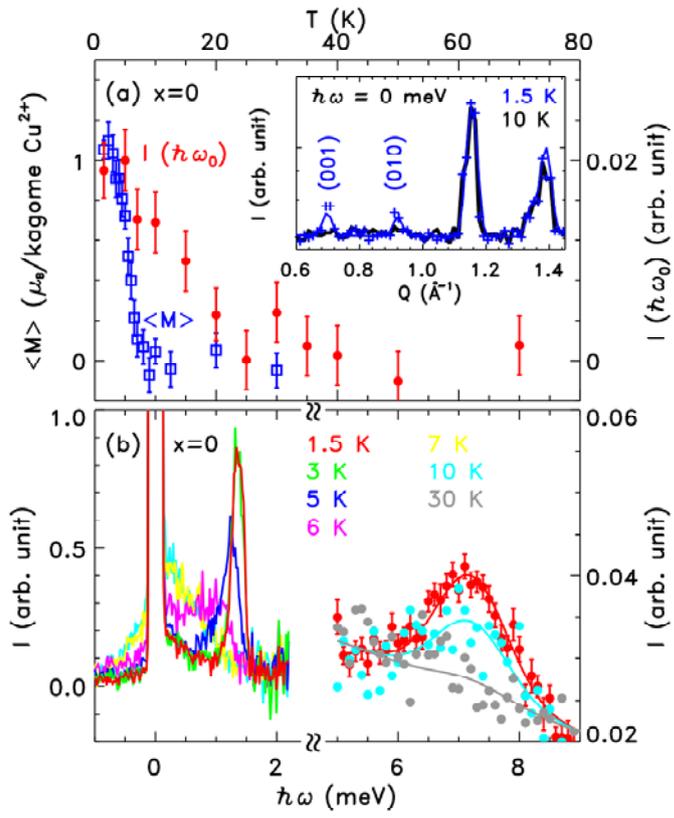

Fig. 2. S.-H. Lee *et al*.

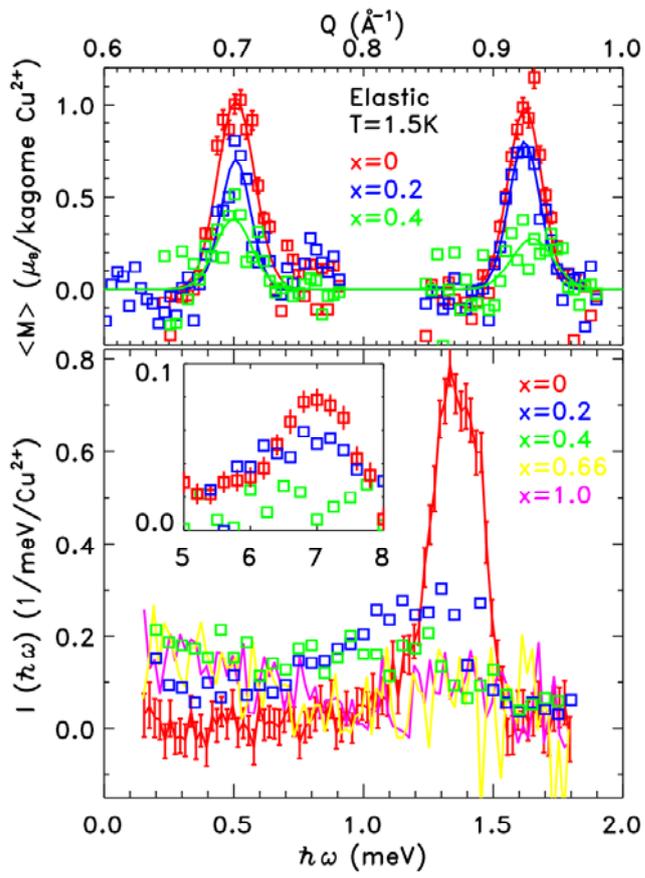

Fig. 3. S.-H. Lee *et al*.

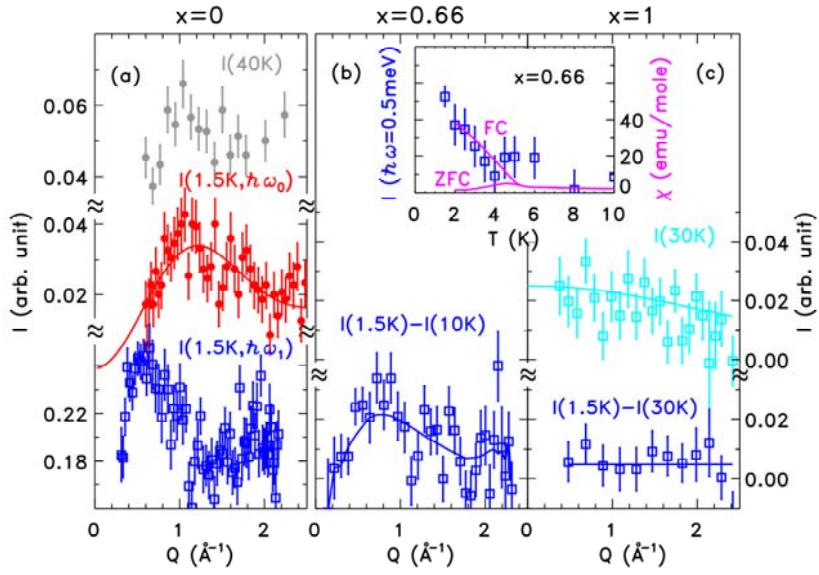

Fig. 4. S.-H. Lee *et al.*

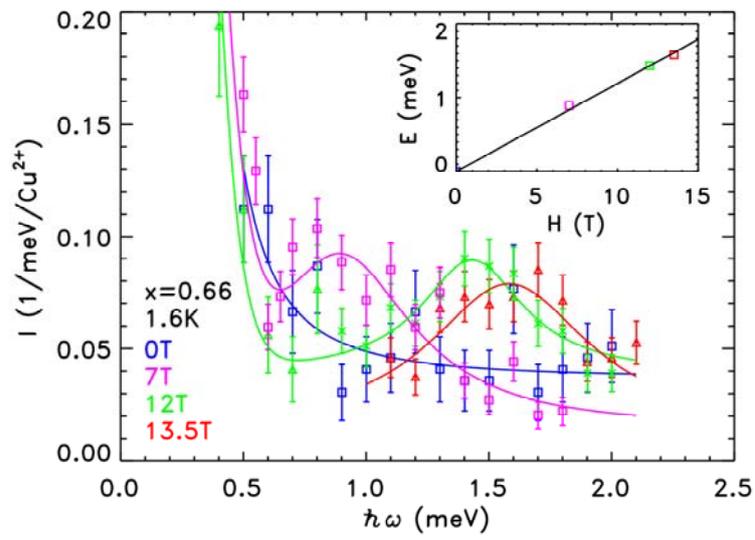

Fig. 5. S.-H. Lee *et al*.